\documentclass[aps,prb,showpacs,twocolumn,floats,epsfig,pdflatex]{revtex4}
\usepackage{amssymb}
\usepackage{amsbsy}
\usepackage{amsmath}
\usepackage{epsfig}
\usepackage{graphicx}
\begin{document}

\title {Superfluid-Insulator transition of two-species bosons with spin-orbit coupling}

\author{Saptarshi Mandal, Kush Saha, and K. Sengupta}

\affiliation {Theoretical Physics Department, Indian Association for
the Cultivation of Science, Kolkata-700032, India. }

\date{\today}

\begin{abstract}

Motivated by recent experiments [Y.J. Lin {\it et al.}, Nature {\bf
471}, 83 (2011)], we study Mott phases and superfluid-insulator (SI)
transitions of two-species ultracold bosonic atoms in a
two-dimensional square optical lattice with nearest neighbor hopping
amplitude $t$ in the presence of a spin-orbit coupling characterized
by a tunable strength $\gamma$. Using both strong-coupling expansion
and Gutzwiller mean-field theory, we chart out the phase diagrams of
the bosons in the presence of such spin-orbit interaction. We
compute the momentum distribution of the bosons in the Mott phase
near the SI transition point and show that it displays precursor
peaks whose position in the Brillouin zone can be varied by tuning
$\gamma$. Our analysis of the critical theory of the transition
unravels the presence of unconventional quantum critical points at
$t/\gamma=0$ which are accompanied by emergence of an additional
gapless mode in the critical region.  We also study the superfluid
phases of the bosons near the SI transition using a Gutzwiller
mean-field theory which reveals the existence of a twisted
superfluid phase with an anisotropic twist angle which depends on
$\gamma$. Finally, we compute the collective modes of the bosons and
point out the presence of reentrant SI transitions as a function of
$\gamma$ for non-zero $t$. We propose experiments to test our
theory.

\end{abstract}


\maketitle

\section{Introduction}
\label{intro}

Ultracold bosons in optical lattices provide us with a wonderful
test bed for studying the physics of strongly correlated bosons in
Mott insulator (MI) and superfluid (SF) phases near the
superfluid-insulator (SI) critical point \cite{Greiner1,Orzel1}. It
is well-known that the low-energy properties of such bosons can be
described by a Bose-Hubbard model which captures the essence of the
SI transition \cite{fisher1,sachdev1,jaksch1}. The analysis of such
a Bose-Hubbard model has been carried out by several authors in the
recent past by using mean-field theory \cite{fisher1,sesh1}, quantum
monte carlo technique \cite{trivedi1,bella1}, projection operator
method \cite{sengupta1}, and strong-coupling expansion
\cite{dupuis1,hrk1}. The advantage of the last method is that it
provides a direct access to boson Green function in the strongly
coupled regime and hence to the momentum distribution of the bosons
in the MI phase near the quantum critical point. In particular, the
method predicts the occurrence of a precursor peak in the momentum
distribution of the bosons in the MI phase near the SI transition
point which has been experimentally verified \cite{spielman1}. More
recently, several theoretical \cite{gaugepapers1} and experimental
\cite{spielman2} proposals of generating artificial Abelian
gauge-fields have been put forth. The strong-coupling expansion has
been also used to describe the SI transition of the bosons in the
presence of such fields \cite{sinha1}; such studies has also been
extended to the case of non-Abelian gauge fields \cite{saha1}.
Further, the method has also been used to study the properties of
the bosons in the presence of a modulated lattice and it has been
shown that such a study can reveal the excitation spectrum of the
bosons both the MI and SF phases near the SI transition point
\cite{sensarma1}.

Spin-orbit coupling plays a key role in shaping the low-energy
properties of several materials including topological insulators
which have been a subject of intense research in recent times
\cite{toprev1}. However, the strength of the spin-orbit coupling is
an intrinsic property of these materials and hence not widely
tunable. More recently, there has been several theoretical proposals
of realization of analogous couplings for neutral bosons in a trap
which has the advantage of generating a tunable spin-orbit coupling
\cite{sopapers1}. One such proposal has recently been realized
experimentally \cite{lin1}. In the experiment of Ref.\
\onlinecite{lin1}, two suitably detuned Raman lasers was used to
generate a momentum and spin-dependent coupling between the $m_F=0$
and $m_F=-1$ hyperfine states of $F=1$ Rb atoms. These two states
acts as two species of the bosons and such a coupling is shown to
generate a term $H_{\rm so} = E_{\ell} k_x \sigma_y/k_{\ell}$ in the
Hamiltonian describing these atoms.  Here $E_{\ell} = \hbar^2
k_{\ell}^2/2m$ is the natural energy unit constructed out of the
wavelength of Raman lasers $k_{\ell}$ and the mass $m$ of the
bosons, and $\vec \sigma$ denotes Pauli matrices in the hyperfine
space ($|m_F=0,-1\rangle$) of the bosons. We note that such a term
is a linear combination of the Rashba $H_{\rm R} \sim (\sigma_x k_y
- \sigma_y k_x)$ and the Dresselhaus $H_{\rm D} \sim -\sigma_x k_y
-\sigma_y k_x$ terms. In addition to the spin-orbit term, the Raman
lasers which are detuned by an energy $\delta $ from the Raman
transition frequency lead to two additional terms in the atom
Hamiltonian. The first of these is directly proportional to the
detuning and is given by $H_{\rm d} = \delta \sigma_y/2$ while the
second term depends on the coupling strength $\Omega$ of the atoms
to the lasers: $H_{\rm c}= \Omega \sigma_z/2$. Together these terms
yield an effective Hamiltonian of the atoms given by
\begin{eqnarray}
H_{\rm eff} = \hbar^2 k^2 I/2m + H_{\rm so} + H_{\rm d} + H_{\rm c}
\label{hambasic}
\end{eqnarray}
where $I$ denotes the identity matrix. We note that the outset that
although the experiments of Ref.\ \onlinecite{lin1} generates
$H_{\rm so}$ which is a linear combination of Rashba and Dresselhaus
terms, there are several theoretical proposals \cite{sopapers1} for
specific generation of either Rashba or Dresselhaus terms using
Raman lasers.

The possibility of realization of spin-orbit coupling for neutral
bosons has led to several theoretical work on the subject
\cite{wu1,yip1,larson1,merkl1,wang1,wu2,zhang1,ss1,vi1,vic1,arun1,wu3}.
Most of these focus on the weak coupling regime (where the boson
interaction can be treated perturbatively) and deal with the nature
of the possible ground states \cite{yip1}, spin-Hall effect in the
presence of a shallow tilted lattice and novel spin excitations
\cite{zhang1,larson1,wu2}, realization of analog of chiral
confinement in one- and multi-dimensional condensates \cite{merkl1},
presence of a spin-stripe phase \cite{wang1}, dynamics of bosons in
the presence of spin-orbit coupling using Gross-Pitaevskii
equations, nature of collective excitations \cite{vi1}, and the
presence of half-quantum vortex excitations \cite{wu1,ss1} of these
bosons in the presence of the spin-orbit term in the SF phase. In
contrast, Refs.\ \onlinecite{vic1,arun1,wu3} focus on the
strong-coupling limit and derive possible effective spin Hamiltonian
to describe these phases for $\Omega=\delta=0$. However, the
analysis of these papers do not provide access to the bosons Green
functions and do not take into account the effect of finite $\delta$
and $\Omega$. One of the central goals of the present work
constitute obtaining such a Green function in the presence of
$\delta$ and $\Omega$ and using it for analyzing the critical theory
of the SI transitions.

In this work we consider two-species bosons in the presence of a
spin-orbit coupling term and in a 2D square optical lattice. The two
species of bosons may be thought to correspond to two hyperfine
states Rb $F=1$ atoms. In the absence of the spin-orbit coupling and
in the presence of the lattice, the Hamiltonian for such a
two-species systems can be written as \cite{issacson1,demler1}
\begin{eqnarray}
H_0 &=& \sum_{i a} [- \mu {\hat n}_{ia} + U  {\hat n}_{ia}( {\hat
n}_{ia}-1)/2] +\lambda U
\sum_i  {\hat n}_{i1}  {\hat n}_{i2} \nonumber\\
&& -  \sum_{\langle ij\rangle a} t_a b_{ia}^{\dagger} b_{ja}
\label{ham0}
\end{eqnarray}
where $b_{ia}$ denotes the bosons annihilation operator on the ${\rm
i^{th}} $ site, $a=1,2$ is the species index, $ {\hat n}_{ia} =
b_{ia}^{\dagger} b_{ia}$ is the boson number operator, $U(\lambda
U)$ is the intra-(inter-)species interaction strength between the
bosons, and $t_a$ (with $t_1=t$ and $t_2= \eta t$) denotes the
nearest neighbor hopping amplitudes. In the presence of the Raman
lasers inducing a Rashba spin-orbit coupling, the additional terms
in the boson Hamiltonian are given, in terms of a two component
boson field $\hat {\Psi}_i= (b_{i1},b_{i2})^{T}$, by
\begin{eqnarray}
H_{1} &=& i \gamma \sum_{\langle ij\rangle} {\hat \Psi}_i^{\dagger}
{\hat z} \cdot \left(
 {\vec \sigma} \times {\vec d}_{ij} \right) {\hat \Psi}_j \nonumber\\
&& + \sum_i  \left[\delta {\hat \Psi}_i^{\dagger} \sigma_y {\hat
\Psi}_i - \Omega {\hat \Psi}_i^{\dagger} \sigma_z {\hat
\Psi}_i\right]. \label{ham1}
\end{eqnarray}
Here the first term represents the lattice analogue of the Rashba
spin-orbit coupling generated by the Raman lasers \cite{haldane1},
${\vec d}_{ij}$ is unit vector along the $x-y$ plane between the
neighboring sites $i$ and $j$, $\Omega$ is the species-dependent
shift in the chemical potential of the bosons, and $\delta$ denotes
the detuning as in Eq.\ \ref{hambasic}. The phase diagram of the
Hamiltonian given by Eq.\ \ref{ham0} has already been studied in
details \cite{issacson1,demler1}; the main purpose of this work is
to study the additional features of the phase diagram due to the
presence of the terms in Eq.\ \ref{ham1}. We note here that for
$\eta=1$, and $\delta=\Omega=0$, $H_0 + H_1$ is formally equivalent
to the Hamiltonian studied in Refs.\ \onlinecite{vic1,arun1,wu3}.

The key results that we obtain from  such a study are the following.
First, we chart out the phase diagram of the bosons in the Mott
phase in the presence of small spin-orbit coupling $\gamma$ and
hopping amplitudes $t_a$. Using a strong coupling theory, we also
obtain the Green function and hence the momentum distribution of the
bosons in these Mott phases. We find that the momentum distribution
of the bosons develops precursor peaks near the SI transition and
show that the position of these peaks in the 2D Brillouin zone can
be continuously tuned from $(k_x,k_y)=(0,0)$ to $(k_x,k_y)= (
\pi/2,\pm \pi/2)$ by varying the relative strengths of the hopping
amplitudes and the spin-orbit coupling $\gamma$. Second, we analyze
the SI transition and show that the transition, for $t_a/\gamma
\simeq 0$, provides an example unconventional quantum critical point
in the sense that it has an additional mode which is gapped in the
superfluid phase but becomes gapless at the critical point. We note
that the presence of such a critical point has been theoretically
conjectured for hardcore bosons with nearest neighbor interactions
\cite{balents1}; however, their presence has not been demonstrated
so far for boson models with finite on-site but no nearest-neighbor
interaction. Third, we chart out the SI phase boundary and study its
variation as a function of $\gamma$ using a Gutzwiller mean-field
theory and show that the ground state in the presence of a finite
$\gamma$ is a twisted superfluid phase and that the twist angle
depends on the ratio $\gamma/t$ \cite{comment1}. Finally, we compute
the collective modes of the bosons and demonstrate that system
undergoes reentrant SI transition which can be accessed by varying
$\gamma$ at a fixed non-zero $t$.

The plan of the rest of the work is as follows. In Sec.\
\ref{miphase}, we chart out the Mott phases of the system and
compute the boson Green function and the momentum distribution in
these phases. This is followed by Sec.\ \ref{crpoint}, where we
construct the effective Landau-Ginzburg (LZ) functionals for such
the SI transitions, and discuss the unconventional nature of the
critical point for $t_a/\gamma \simeq 0$. In Sec.\ \ref{mfta}, we
use Gutzwiller mean-field theory to chart out the
superfluid-insulator phase boundary and show that the superfluid
ground state is a twisted superfluid. This is followed by Sec.\
\ref{goldstone} where we use the LZ functionals constructed in Sec.\
\ref{crpoint} to compute the collective modes of the bosons in the
superfluid phases near the SI transition. Finally, we present a
discussion of the work and conclude in Sec.\ \ref{conc}.

\section {Mott phases and the boson momentum distribution}
\label{miphase}

\subsection {Mott phase in the atomic limit}
\label{miphasea}

In this Section, we shall chart out the Mott phases of the system in
the so-called Mott or atomic limit where $\gamma = t_a=0$. The
Hamiltonian of the system in this limit is given by
\begin{eqnarray}
H_{\rm Mott} &=& \sum_{i a} [- [\mu + \Omega {\rm Sgn}(a)] {\hat
n}_{ia} + U
{\hat n}_{ia}( {\hat n}_{ia}-1)/2] \nonumber\\
&& + \sum_i \lambda U  {\hat n}_{i1} {\hat n}_{i2} + i \delta
\left(b_{i2}^{\dagger} b_{i1} - b_{i1}^{\dagger} b_{i2} \right),
\label{mottham1}
\end{eqnarray}
where ${\rm Sgn}(a) =\pm 1$ for $a=1,2$. Since all the terms in the
Hamiltonian are on-site, one can choose a Gutzwiller like
wavefunction $|\psi_{MI} \rangle = \prod_{\bf r} f_{n_1 n_2} |n_1,
n_2\rangle$, where $n_{1(2)}$ denotes the occupation of bosons of
species $1(2)$ at a lattice site ${\bf r}$, and compute the energy
of the system $ E[\{f_{n_1 n_2} \}] = \langle \psi_{MI}| H_{\rm
mott} | \psi_{MI}\rangle$. Further, since the total number of
particles per site $n= n_1+n_2$ commutes with $H_{\rm Mott}$, the
Hamiltonian decomposes into different sectors labeled by $n$. Thus,
one can separately compute and compare the energy functionals $E_n
\equiv E_n[\{f_{n_1 n_2}\}]$ for each $n$ to find the ground state.
For $n=0$, $E_0=0$ while for $n=1$ the energy functionals reads
\begin{eqnarray}
E_1 &=& -(\mu +\Omega) |f_{1 0}|^2 + (\Omega-\mu) |f_{0 1}|^2
\nonumber\\
&& +i \delta ( f_{01}^{\ast} f_{10} - f_{10}^{\ast} f_{01}).
\label{en1}
\end{eqnarray}
A similar expression for $E_2$ and $E_3$ can also be written down.
For $n=2$, we find
\begin{eqnarray}
&& E_2= \Psi^{\ast}_2\left(
\begin{array}{ccc} -2\mu+\lambda U & i \sqrt{2}\delta& -i\sqrt{2}\delta
\\ -i \sqrt{2}\delta&-2\mu - 2\Omega + U&0 \\
i\sqrt{2}\delta&0&-2\mu+2\Omega+ U \end{array} \right) \Psi_2
\label{en2} \nonumber\\
\end{eqnarray}
where $\Psi_2 = (f_{11},f_{20},f_{02})^T$. Similarly for $n=3$, one
can define $\Psi_3 = (f_{12},f_{21},f_{03},f_{30})^T$ and obtain
\begin{widetext}
\begin{eqnarray}
&& E_3= \Psi^{\ast}_3 \left( \begin{array}{cccc} -3\mu+2\lambda
U+\Omega+ U & i 2\delta& -i\sqrt{3}\delta&0 \\ -i
2\delta&-3\mu - \Omega +2\lambda U + U &0&i\sqrt{3}\delta \\
i\sqrt{3}\delta&0&-3\mu+3\Omega+3U & 0\\
0&-i\sqrt{3}\delta&0&-3\mu-3\Omega +3U \end{array} \right) \Psi_3.
\label{en3}
\end{eqnarray}
\end{widetext}
The ground state of the system is then determined by minimizing
$E_n$ for a given set of dimensionless parameters $\mu/U, \,
\lambda, \, \Omega/U, \, {\rm and} \, \delta/U$.

To chart out the phase diagram, we first consider case $\delta=0$.
In this case, all the off-diagonal terms in Eq.\ \ref{en2} and
\ref{en3} vanish and one obtains
\begin{eqnarray}
&& \mathcal{E}_{10}=-\mu -\Omega,~~\mathcal{E}_{01}=-\mu+ \Omega,~~ \mathcal{E}_{11}=- 2\mu + \lambda U, \nonumber \\
&& \mathcal{E}_{20}=- 2\mu -2 \Omega+ U,~~\mathcal{E}_{02}=- 2\mu +2 \Omega+U, \nonumber\\
&& \mathcal{E}_{30}=-3\mu-3\Omega+3U,~~
\mathcal{E}_{03}=-3\mu+3\Omega+3U, \nonumber\\
&&\mathcal{E}_{21}=-3\mu+2\lambda U -\Omega+ U,\nonumber\\
&&\mathcal{E}_{12}=-3\mu+2\lambda U +\Omega+ U. \label{del0en}
\end{eqnarray}

\begin{figure}
\includegraphics[height=70mm,width=75mm]{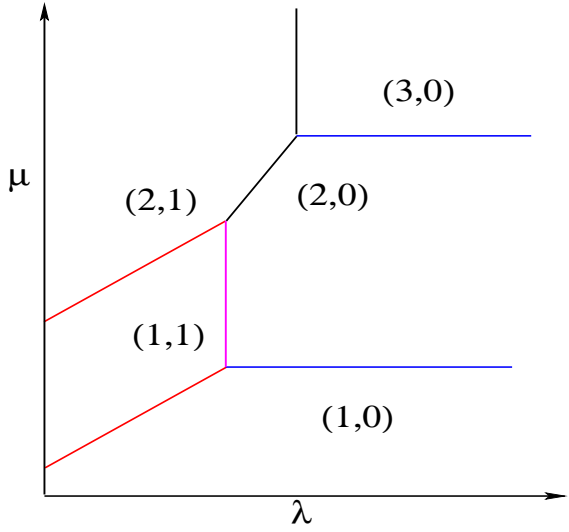}
\caption{ Schematic phase-diagram showing preferred particle
distribution in $\mu$-$\lambda$ plane for $\delta=0$. The pairs of
numbers $(n_1,n_2)$ denotes the particle numbers of the two species
at each site.} \label{fig1}
\end{figure}

The MI phase diagram for $\delta=0$ is shown in Fig.\ \ref{fig1}. We
note from Eq.\ \ref{del0en} that the boundary between MI phase
$(1,0)$ and $(1,1)$ is determined by $\mathcal{E}_{10} =
\mathcal{E}_{11}$ leading to the condition $\mu = \Omega + \lambda
U$. Similarly, the boundary between $(1,0)$ and $(2,0)$ phases is
determined by the condition $\mu = -\Omega +U$ while that between
the $(1,1)$ and $(2,0)$ phases is given by $\lambda U = -2 \Omega+
U$.

\begin{figure}
\includegraphics[width=\linewidth]{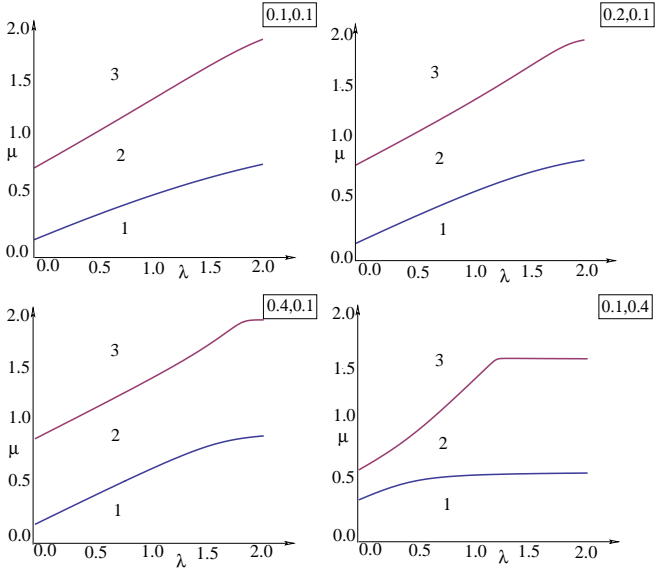}
\caption{ Schematic phase-diagram of the MI phase showing in the
$\mu$-$\lambda$ plane for finite $\Omega$ and $\delta$. The numbers
$1$, $2$, and $3$ denotes the total boson number $n$ at a given
regime. The pair of in the box for each panel denote the values of
$(\Omega, \delta)$ for which the phase diagram has been
drawn.}\label{fig2}
\end{figure}

For finite $\delta$, the energy of different Mott phases are
determined by Eqs.\ \ref{en1}, \ref{en2} and \ref{en3}. Using these
equations, we find the ground state numerically as function of $\mu$
and $\lambda$ for several representative values of $\Omega$ and
$\delta$ as shown in Fig.\ \ref{fig2}. We note that the main effect
of $\delta$ is to smoothen out the phase boundary between the phases
and to realize a MI ground which a linear superposition of states
with different $n_1$ and $n_2$ with a fixed $n=n_1+n_2$. For
example, the ground state with $n=1$ in Fig.\ \ref{fig2} is a linear
superposition of the states $(1,0)$ and $(0,1)$. The overlap of a
state $(n_1, n_2)$ with the ground state with $n=n_1+n_2$ depends on
the precise values of $\delta$ and $\Omega$.

\subsection{Momentum distribution in the MI phase}
\label{miphaseb}

In this section, we shall compute the momentum distribution of the
Green function in the Mott phase for which $n=1$. The calculations
can be generalized to any $n$ in a straightforward manner; however
this requires handling quite complicated algebra which we refrain
from in this work.

First, let us consider the Green function of the bosons in the MI
phase in the atomic limit. For $n=1$, the Green function is a $2
\times 2$ matrix given by
\begin{eqnarray}
&&G_0(\tau,\tau^{\prime})=
\left( \begin{array}{cc}  \langle \mathcal{T}b^{\dagger}_1(\tau) b_1(\tau^{\prime})
\rangle &  \langle \mathcal{T}b^{\dagger}_1(\tau) b_2(\tau^{\prime}) \rangle\\
\langle \mathcal{T} b^{\dagger}_2(\tau) b_1(\tau^{\prime}) \rangle &
\langle  \mathcal{T}b^{\dagger}_2(\tau) b_2(\tau^{\prime}) \rangle
\end{array} \right) \label{gf1}
\end{eqnarray}
To compute the Green function, we first consider the eigenenergies
of $H_{\rm Mott}$. These are obtained by diagonalizing $E_n$ for the
$n$ particle sector; for computing the zero-temperature Green
function for the $n=1$, sector, we shall need the expressions of
these energies for $n=1$ and $n=2$ sectors. For $n=1$, let us denote
these energies by $\mathcal{E}^1_1$ and $\mathcal{E}^1_2$ with
$\mathcal{E}^1_1 < \mathcal{E}^1_2$. it can be easily seen from Eq.\
\ref{en1}, that the corresponding eigenstates
$|\mathcal{E}^1_1\rangle$ and $|\mathcal{E}^1_2\rangle$ are related
to the states $|1,0\rangle$ and $|0,1\rangle$ by
\begin{eqnarray}
\left(\begin{array}{c}
|\mathcal{E}^1_1 \rangle \\
|\mathcal{E}^1_2\rangle \end{array} \right) =\left( \begin{array}{cc} u_1&v_1 \\
u_2&v_2 \end{array} \right) \left(\begin{array}{c}
|1,0 \rangle \\
|0,1\rangle \end{array} \right) \label{uvrel}
\end{eqnarray}
where $v_1= u_2= \delta/{\mathcal D}$, and $u_1=v_2 = -i (\Omega
+\sqrt{\delta^2 + \Omega^2})/{\mathcal D}$ where ${\mathcal D}=
[\delta^2 +(\Omega^2 +\sqrt{\delta^2+\Omega^2})]^{1/2}$. Similarly
for the $n=2$ sectors, we denote the eigenenergies and corresponding
eigenfunctions of $H_{\rm Mott}$ by ${\mathcal E}_{1,2,3}^2$ and
$|{\mathcal E}_{1,2,3}^2\rangle$ respectively. From Eq.\ \ref{en2},
we find that the states $|{\mathcal E}_{1,2,3}^2\rangle$ are related
to $|1,1\rangle$, $|2,0\rangle$, and $|0,2\rangle$ by
\begin{eqnarray}
\left(\begin{array}{c}
|1,1 \rangle \\
|2,0\rangle \\
|0,2 \rangle \end{array} \right)=\left( \begin{array}{ccc} x_1&y_1&z_1 \\
x_2&y_2&z_2\\
x_3&y_3&z_3 \end{array} \right) \left(\begin{array}{c}
|\mathcal{E}^2_1 \rangle \\
|\mathcal{E}^2_2\rangle \\ |\mathcal{E}^2_3 \rangle \end{array}
\right)
\end{eqnarray}
where the expressions of $x_i$, $y_i$ and $z_i$ can be found by
diagonalizing the energy functional $E_2$ (Eq.\ \ref{en2}). These
coefficients are found numerically in the present work for finite
$\delta$. Here we note that $x_i$, $y_i$ and $z_i$ are imaginary for
$i=2,3$ and real for $i=1$. Using these expressions, a
straightforward calculation following Ref.\ \onlinecite{dupuis1}
yields the atomic limit Green functions as
\begin{eqnarray}
&&G^0_{11}(iw)= -\frac{|u_1|^2}{{\mathcal E}^1_1 -iw} + \sum^3_{j=1}
\frac{T^{j}_{11}}{\mathcal{E}^2_j -{\mathcal E}^1_1 -iw},\nonumber\\
&&G^0_{12}(iw)= -\frac{u^*_1v_1}{{\mathcal E}^1_1 -iw} +
\sum^3_{j=1} \frac{T^{j}_{12}}{\mathcal{E}^2_j -{\mathcal E}^1_1 -iw},\nonumber\\
&&G^0_{21}(iw)= -\frac{u_1v^*_1}{{\mathcal E}^1_1 -iw} +
\sum^3_{j=1} \frac{T^{j}_{21}}{\mathcal{E}^2_j -{\mathcal E}^1_1 -iw},\nonumber\\
&&G^0_{22}(iw)= - \frac{|v_1|^2}{{\mathcal E}^1_1 -iw} +
\sum^3_{j=1} \frac{T^{j}_{22}}{\mathcal{E}^2_j -{\mathcal E}^1_1
-iw}, \label{gfmott}
\end{eqnarray}
where $i\omega$ denotes Matsubara frequency and $T^j_{ab}$, for
$j=1,2,3$ and $a,b=1,2$ are given by
\begin{eqnarray}
T^{1}_{11} &=&  2|u_1|^2 |x_2|^2 + \sqrt{2}u_1v^*_1x^*_1x_2+
\sqrt{2}u^*_1v_1x_1x^*_2 \nonumber\\
&& + |v_1|^2 |x_1|^2, \nonumber\\
T^{1}_{22} &=& |u_1|^2 |x_1|^2 + \sqrt{2}u_1v^*_1x^*_1x_3+
\sqrt{2}u^*_1v_1x_1x^*_3\nonumber\\
&& +2 |v_1|^2 |x_3|^2, \nonumber\\
T^{1}_{21}&=& T_{12}^{1 \ast} = \sqrt{2} |u_1|^2 x^*_2x_1 +
u_1v^*_1|x_1|^2 \nonumber\\
&& + \sqrt{2}|v_1|^2x^*_1x_3 + 2 v^*_1u_1 x^*_2x_3, \label{texpmott}
\end{eqnarray}
and $T_{ab}^2$ and $T_{ab}^3$ are obtained by replacing all $x_i$s
in the expression of $T_{ab}^1$ by $y_i$ and $z_i$ respectively.
Note that, when analytically continued to real frequencies using the
prescription $i \omega \to \omega +  i\epsilon$, $G^0_{ij}(\omega)$
is imaginary for $i\ne j$ and real for $i=j$ for $\epsilon=0$.

The Green functions obtained in Eq.\ \ref{gfmott} can be easily
understood as follows. Each term $G^0_{ab}$ receives contribution
from a hole branch which corresponds to removal of one particle from
the Mott state which cost an energy ${\mathcal E}_1^1$ in the atomic
limit. The other terms represents contribution from the different
possible particle branches which corresponds to addition of a
particle over the ground state with $n=1$ and cost energies
$\mathcal{E}^2_j -\mathcal {E}^1_1$ for $j=1,2,3$. The poles of the
Green functions occur at these particle and hole excitation
energies.

To obtain the Green function for finite nearest-neighbor terms $t_a$
and $\gamma$, we follow the procedure introduced in Ref.\
\onlinecite{dupuis1}. First, we define the bosonic fields as
$\psi_{a i}(\tau) \equiv \psi_a({\bf r_i},\tau)$, where $a=1,2$, $i$
denote the site index of the optical lattice and $\tau$ is the
imaginary time. In terms of these fields, the nearest-neighbor
hopping and spin-orbit coupling terms given by Eqs.\ \ref{ham0} and
\ref{ham1} can be written as
\begin{widetext}
\begin{eqnarray}
S_0 &=& \int_0^{\beta}  d\tau \sum_{\langle ij \rangle} \Big[
\left(\psi^{\ast}_{1i} \psi_{2i}^{\ast} \right) \Lambda
\left(\begin{array}{c}
\psi_{1j} \\
\psi_{2j} \end{array} \right)+{\rm h.c} \Big]  \nonumber\\
\Lambda &=& \left( \begin{array}{cc} -t_1& i\gamma (\delta_{j,i\pm{\hat y}} +(-1)^a i \delta_{j,i\pm{\hat x}})  \\
-i\gamma (\delta_{j,i\pm{\hat y}} - (-1)^a i \delta_{j,i\pm{\hat
x}})&-t_2 \end{array} \right), \label{keterm1}
\end{eqnarray}
\end{widetext}
where we have omitted the $\tau$ index of the boson fields for
clarity, $\beta = 1/k_B T$ is the inverse temperature and $k_B$ is
the Boltzman constant which will be subsequently set to unity. We
then write down the coherent state path integral for the bosons and
decouple the nearest-neighbor hopping and spin-orbit coupling terms
by two Hubbard-Stratonovitch fields $\Delta_{i}(\tau)=
[\Delta_{1i}(\tau) \Delta_{2i}(\tau)]^T$ and so that the partition
function of the bosons can be written as (with $\hbar=1$)
\begin{eqnarray}
Z &=& \int {\mathcal D} \psi_a^{\ast} {\mathcal D} \psi_a {\mathcal
D} \Delta_a  e^{-S_1[\psi_a^{\ast}, \psi_a,
\Delta_a]} \nonumber\\
S_1 &=& \int_0^{\beta}  d\tau \Big [ \sum_{ia} \Big(\psi_{ia}
\partial_{\tau} \psi_{ia} - (\Delta_i^{\ast} \psi_{ia} +{\rm h.c}) \Big)
+ H_{\rm Mott} \nonumber\\
&& - \sum_{\langle ij\rangle} \Delta_i^{\ast} \Lambda^{-1} \Delta_j
\Big] \label{hs1}
\end{eqnarray}
Next, we introduce a second Hubbard-Stratonovitch field
$\Phi_i(\tau) = [\Phi_{1i}(\tau), \Phi_{2i}(\tau)]^T$ to decouple
the last term in $S_1$ (Eq.\ \ref{hs1}). This leads to
\begin{eqnarray}
Z &=& \int {\mathcal D} \psi_a^{\ast} {\mathcal D} \psi_a {\mathcal
D} \Delta_a {\mathcal D} \Phi_a e^{-S_2[\psi_a^{\ast}, \psi_a,
\Delta_a, \Phi_a]} \nonumber\\
S_2 &=& \int_0^{\infty}  d\tau \Big [ \sum_{ia} \Big(\psi_{ia}
\partial_{\tau} \psi_{ia} + [\Delta_i^{\ast}(\Phi_{ia} - \psi_{ia}) +{\rm h.c} \Big) \nonumber\\
&& + H_{\rm Mott} - \sum_{\langle ij\rangle} \Phi_i^{\ast} \Lambda
\Phi_j \Big] \label{hs2}
\end{eqnarray}
We note that the field $\Phi_{ia}(\tau)$ have exactly the same
correlators as the original boson fields \cite{dupuis1}. With this
observation, we integrate out the fields $\Delta_{ia}(\tau)$ and
$\psi_{ia}(\tau)$ to obtain an effective action in terms of the
field $\Phi_{ia}(\tau)$. The details of the procedure for doing so
is elaborated in Ref.\ \onlinecite{dupuis1}. After some algebra, the
quadratic and the quartic part of the resultant action is obtained
to be
\begin{eqnarray}
S^{(2)}_{\rm eff} &=&  \frac{1}{\beta} \sum_{ \omega_n;a,b} \int
\frac{d^2k}{(2\pi)^2} \Phi_a^{\ast} (k) \left[ - G^{0-1} (\omega_n)
\right. \nonumber\\
&& \left. + \Lambda({\bf k}) \right]_{ab} \Phi_b(k) \label{quadac}
\\
S^{(4)}_{\rm eff} &=& \frac{g}{2} \int_0^{\beta} d\tau \int d^2 x
\left|\sum_a \Phi_a^{\ast}(r) \Phi_a(r)\right|^2 \label{quarac}
\end{eqnarray}
where $G^0$ denotes the boson Green functions in the atomic limit,
$k \equiv (\omega_n;{\bf k})$, $r \equiv (\tau;{\bf x})$, and
$\Lambda({\bf k})$ is given by
\begin{eqnarray}
\Lambda({\bf k}) &=& -2 \left(\begin{array}{cc} t_1(\cos k_x
+ \cos k_y)&  \gamma (i\sin k_x + \sin k_y) \\
\gamma (-i \sin k_x + \sin k_y)& t_2(\cos k_x + \cos k_y)
\end{array} \right).  \nonumber\\  \label{endisp1}
\end{eqnarray}
In what follows we shall analyze $S^{(2)}_{\rm eff}$ and
$S^{(4)}_{\rm eff}$ within mean-field theory to obtain the
properties of MI and SF phases of the bosons. We shall neglect all
high order terms in the boson effective action which can be shown to
be irrelevant in the low-energy, low-momentum limit \cite{dupuis1}.

\begin{figure}
\begin{center}
\includegraphics[width=0.9 \linewidth]{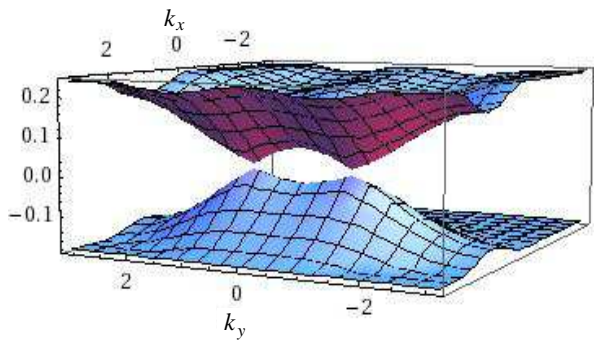}
\caption{ The highest hole and the lowest particle bands of the
bosons for $\lambda=0.4, \, \Omega=0.01 U, \, \delta=0.005 U, \,
t=0.02 U, \, \gamma=0.049 U, \, \mu=0.175 U, \, {\rm and}\,
\eta=0.5$.} \label{fig3}
\end{center}
\end{figure}

In the MI phase, $\langle \Phi_{a}(k) \rangle =0$ and the boson
action, within mean-field theory, is given by $S_{\rm eff}^{(2)}$.
The momentum distribution of the bosons in the MI phase at zero
temperature can then obtained from the boson Green function $G_{\rm
eff}(k) = \left[- G^{0-1} (\omega) + \Lambda({\bf k}) \right]^{-1}$
as
\begin{eqnarray}
n({\bf k}) &=& \int_{-\infty}^0  \frac{d \omega}{2 \pi} {\rm Tr}\,
G_{\rm eff} (k) \label{momdist1}
\end{eqnarray}
where ${\rm Tr}$ denotes matrix trace and we have used $\omega_n
\to\omega + i\epsilon$ for analytic continuation to real
frequencies. To evaluate the integral, we note that the integrand
${\rm Tr}\, G_{\rm eff}(k)$ is invariant under an unitary
transformation; consequently $n({\bf k})$ in Eq.\ \ref{momdist1} can
be written as $n({\bf k}) = \int_{-\infty}^0 d\omega {\rm Tr}\,
G^d_{\rm eff}(k) /(2\pi)$ where $G^d_{\rm eff}(k)$ is obtained by
diagonalizing $G_{\rm eff}(k)$ via an unitary transformation and can
be written as
\begin{eqnarray}
G_{\rm eff}^d (k) = \prod_{i=1}^{n_{band}} \frac{{\mathcal
A}_i(\omega;{\bf k})}{\left[\omega - \epsilon_i({\bf k}) \right]},
\label{gdiag}
\end{eqnarray}
where $\epsilon_i({\bf k})$ are the band energies which are obtained
as solution of ${\rm Det} [G^{-1}_{\rm eff}(\epsilon;{\bf k})] =0$,
${\mathcal A}_i(\omega,{\bf k})$ are the residue of the Green
function at the pole $\omega=\epsilon_i({\bf k})$ which has to be
determined numerically for finite $\delta$, and $n_{band}$ is the
total number of such bands. The equation for determining these bands
can be written using Eq.\ \ref{gfmott}, \ref{endisp1}, and
\ref{quadac} as
\begin{eqnarray}
&& \{[G_0^{-1}(\epsilon)]_{11} + 2t_1 (\cos(k_x)+\cos(k_y))\}
\nonumber\\
&& \times \{ [G_0^{-1}(\epsilon)]_{22} + 2t_2
(\cos(k_x)+\cos(k_y))\}  \nonumber\\
&& = 4\sin^2(k_y)\gamma^2+ \left[2\gamma \sin(k_x) +
[G_0^{'-1}(\epsilon)]_{12}\right]^2 \label{bandisp1}
\end{eqnarray}
where we have used the fact that $[G_0^{-1}(\epsilon)]_{12}=
[G_0^{-1}(\epsilon)]_{21}^{\ast}$ and $[G_0^{'-1}(\epsilon)]_{12}=-i
[G_0^{-1}(\epsilon)]_{12}$. We note that for finite $\delta$,
$[G_0^{-1}(\epsilon)]_{12} \ne 0$. Consequently, Eq.\ \ref{bandisp1}
is invariant under $k_y \to -k_y$ but not under $k_x \to -k_x$; thus
the energy bands satisfy $\epsilon_i(k_x,k_y)= \epsilon_i (k_x,-k_y)
\ne \epsilon_i(-k_x, k_y)$. A plot of the highest negative and the
lowest positive energy bands for representative values of parameters
is shown in Fig.\ \ref{fig3}. The plot clearly indicates two minima
at $k_y=\pm k_y^0=\pm k_0$ and $k_x= k_0$. Also, we find that for
the above-mentioned parameter values $n_{band}=8$; there are two
bands with negative and six bands with positive energies.

To compute $n({\bf k})$, we note that the contribution to $n({\bf
k})$ comes from the bands for which $\epsilon({\bf k}) \le 0$.
Labeling such energy bands as $\epsilon^{-}_i({\bf k})$ and denoting
their number by $n_{band}^-$, one obtains the momentum distribution
as
\begin{eqnarray}
n({\bf k}) &=& \sum_{i=1}^{n^-_{band}} {\mathcal
A}_i(\epsilon_i({\bf k});{\bf k})\prod_{j=1}^{n_{band}}
\frac{(1-\delta_{ij}){\mathcal A}_j(\epsilon_i({\bf k}); {\bf
k})}{[\epsilon_j({\bf k})-\epsilon_i^-({\bf k})]}
\label{momdistfinal} \nonumber\\
\end{eqnarray}
where the sum extends over all bands with $\epsilon_i \le 0$.

\begin{figure}
\begin{center}
\includegraphics[width=0.9 \linewidth]{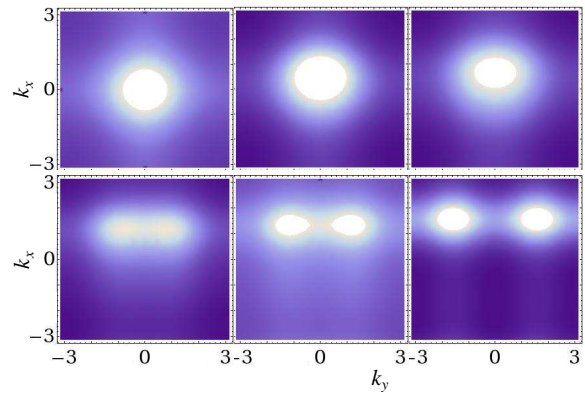}
\caption{ The momentum distribution of the bosons in the MI phase
showing the precursor peaks moving from center of the Brillouin zone
to $(\pi/2,\pm \pi/2)$ with increasing $\gamma/t$. The plots
correspond to $\mu=0.2 U$ and have (from top left to bottom right)
$(t, \gamma)$ to be $(0.03,0)$, $(0.03, 0.02)$, $(0.025,0.025)$,
$(0.015,0.04)$, $(0.01,0.045)$, and $(0.0,0.048)$ in units of $U$.
The lighter colors indicate larger values of $n({\bf k})$. All other
parameter values are same in Fig \ref{fig3}.} \label{fig4}
\end{center}
\end{figure}

The plot of $n({\bf k})$ is shown in Fig.\ \ref{fig4}. As expected,
we find $n({\bf k})$ develops precursor peaks as one approaches the
SI transition point by increasing $t_a$ and $\gamma$. This feature
of $n({\bf k})$ can be easily understood from Eq.\
\ref{momdistfinal} and Fig.\ \ref{fig3} by noting the following
points. First, the energy bands are independent of ${\bf k}$ for
$t_a=\gamma=0$ (atomic limit) leading to a flat $n({\bf k})$.
Second, as we approach the SI transition, the gap between the
highest band with $\epsilon_i({\bf k}) < 0$ and the lowest band with
$\epsilon_i({\bf k}) > 0$ decreases at special points $(k_x^0, \pm
k_y^0)$ in the Brillouin zone. This results in peaks of $n({\bf k})$
at these points as we approach the SI transition. These peaks are
precursors to the SI transition at which the bands touch; the
position of these precursor peaks depend on the ratio $\gamma/t_1$
(for a fixed $\eta$) and can be continually tuned from $(\pi/2,\pm
\pi/2)$ to $(0,0)$ as $\gamma/t_1$ is decreased. Note that since
$\epsilon_i(k_x,k_y) = \epsilon_i(k_x,-k_y)$, both $(k_x^0,k_y^0)$
and $(k_x^0,-k_y^0)$ correspond to the peak position; however, since
$\epsilon_i(k_x,k_y) \ne \epsilon_i(-k_x,k_y)$ for finite $\delta$,
$n({\bf k})$ need not (and does not) have a peak at $(-k_x^0,k_y^0)$
unless $\delta=0$. Numerically, we find $k_x^0=k_y^0= k_0$ for all
the parameter range we study. Thus our work demonstrates that the
key effect of the spin-orbit coupling is to shift these precursor
peaks from $(0,0)$ to finite momenta $(k_0,\pm k_0)$ in the
Brillouin zone. In the next section, we shall investigate the effect
of this shift on the SI transition point.

\section {Superfluid-Insulator transition}
\label{crpoint}

In this section, we shall analyze the SI transition for two species
bosons with spin-orbit coupling. We use the strong coupling Green
function developed in Sec.\ \ref{miphaseb} to construct an effective
low-energy critical theory for the transition. This is followed by
the analysis of the critical theory in Sec.\ \ref{crth}.

\subsection{Critical Theory}
\label{trsc}

In this section, we analyze the critical theory of the
superfluid-insulator transition using $S^{(2)}_{\rm eff}$ and
$S^{(4)}_{\rm eff}$ (Eqs.\ \ref{quadac} and \ref{quarac}) derived in
Sec.\ \ref{miphaseb}. These terms provide the microscopic basis for
construction of an effective Landau-Ginzburg functional for the
MI-SF transition. The analytical calculations in this section will
be carried out for $\delta=0$ for simplicity; however, we shall
provide qualitative statements for $\delta \ne 0$ case at the end of
this section.

We consider approaching the critical point from the MI side. For
$\delta=0$, the on-site Green function $G^0$ is diagonal with the
elements $G_{11}^0$ and $G_{22}^0$ given by
\begin{eqnarray}
G_{11}^0 (\omega) &=& \frac{-1}{\omega +E_0} +  \frac{2}{\omega +E_0
-U}, \quad G_{22}^0 (\omega) = \frac{1}{\omega - E_1}, \nonumber\\
E_0 &=& \mu +\Omega, \quad E_1 = \mu - \Omega - \lambda U.
\label{gratd0}
\end{eqnarray}
This allows us to write $(G^0)^{-1}$ as a diagonal matrix
\begin{eqnarray}
(G^0)^{-1} &=& \left( \begin{array}{cc} F_1(\omega) & 0 \\
0 & F_2(\omega) \end{array} \right)  \label{gatinvd0}  \\
F_1(\omega) &=& \omega +E_0 - \frac{2U(\omega+E_0)}{\omega +E_0 +
U}, \quad F_2(\omega) = \omega -E_1. \nonumber
\end{eqnarray}
Using Eq.\ \ref{gatinvd0}, one can write the effective action
$S^{(2)}_{\rm eff}(\delta=0)$ as
\begin{widetext}
\begin{eqnarray}
S^{(2)}_{\rm eff}(\delta=0) &=& - \sum_{a,b=1,2} \int \frac{d^2k d
\omega}{(2\pi)^3} \Phi_a^{\ast} (k) G^{-1}(k) \Phi_b(k)
\nonumber\\
G^{-1}(k) &=& \left( \begin{array}{cc} F_1(\omega)
+2t[\cos(k_x)+\cos(k_y)] & 2\gamma [i \sin(k_x) + \sin(k_y)]  \\ 2
\gamma [-i \sin(k_x)+ \sin(k_y)] & F_2(\omega) + 2t
[\cos(k_x)+\cos(k_y)] \end{array} \right)  \label{ginvd0}
\end{eqnarray}
\end{widetext}
where $k\equiv (\omega, {\bf k})$. Diagonalizing $G^{-1}(k)$, we
find the two eigenvalues to be
\begin{eqnarray}
\lambda_{\pm} &=& \frac{1}{2} \Big [ F_+(\omega) + 4t
[\cos(k_x)+\cos(k_y)]
\nonumber\\
&& \pm \sqrt{ F_-^2(\omega) + 16 \gamma^2 [\sin^2(k_x) +\sin^2
(k_y)]} \Big]
\end{eqnarray}
where $F_{\pm}(\omega) = F_1 (\omega)\pm F_2(\omega)$. Thus the
quadratic part of the effective action of the bosons can be written
as
\begin{eqnarray}
S^{(2)}_{\rm eff}(\delta=0) &=& - \sum_{a=\pm} \int \frac{d^2k d
\omega}{(2\pi)^3} \Phi_a^{\ast} (k) \lambda_a(k) \Phi_a(k)
\label{gdiagd0}
\end{eqnarray}
where $\Phi_{+(-)} = \alpha_1^{+(-)} \Phi_1 + \alpha_2^{+(-)}
\Phi_2$ are linear combinations of the fields $\Phi_1$ and $\Phi_2$
and $\alpha_{1,2}^{\pm}$ are the components of eigenfunctions of
$G^{-1}(k)$ corresponding to eigenvalues of $\lambda_{\pm}$ given by
\begin{eqnarray}
\frac{\alpha_2^{\pm}}{\alpha_1^{\pm}} &=& - \frac{ F_1(\omega) +
2t[\cos(k_x)+\cos(k_y)] -\lambda_{\pm}(k)}{2\gamma [i \sin(k_x)+
\sin(k_y)]} \label{eigend0}
\end{eqnarray}

\begin{figure}
\includegraphics[width=0.5\textwidth]{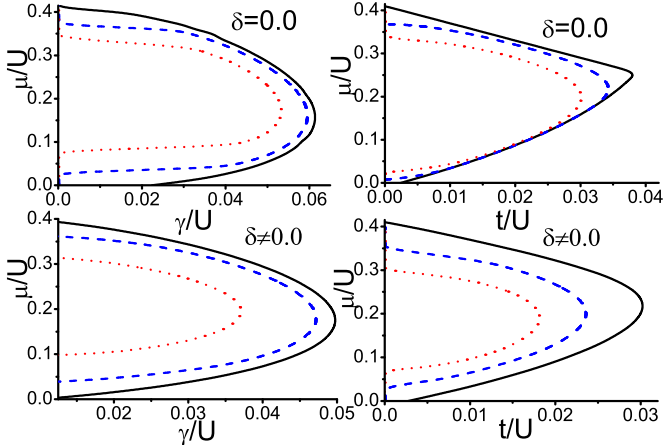}
\caption{The phase boundary in the $\mu-\gamma$ (left panels) and
$\mu-t$ plane (right panels) for $\eta=1$, and $\Omega=0.01 U$ as
obtained from the strong-coupling analysis. The top panels have
$\delta=0$ while the bottom panels have $\delta=0.005 U$. The values
of $t$ for the left panels corresponding to different lines are
$t/U=0.0$ (black solid), $t/U=0.01$ (blue dashed), $0.02$ (red
dotted). For the right panels, $\gamma/U=0$ (black solid line),
$0.03$ (blue dashed line) and $0.04$ (red dotted line).
}\label{fig6}
\end{figure}

\begin{figure}
\includegraphics[width=\linewidth]{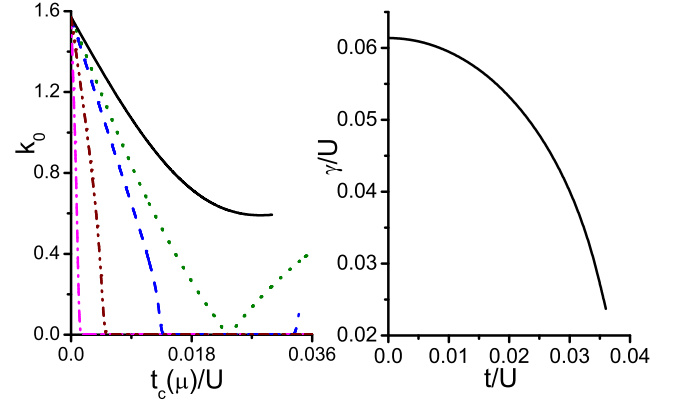}
\caption{Left panel: Plot of $k_0$ against $t_c(\mu)$ for
$\gamma/U=0.01$(pink dash dot dotted), $0.02$ (yellow dash dotted ),
$0.03$ (blue dotted), $0.0333$ (green dashed) and $0.04$(black
solid) line. Right panel: Plot of the line with $z=1$ quantum phase
transition in the $\gamma-t$ plane as obtained from solution of
$K_0=0$.For all plots $\delta=0$, $\lambda=0.4$, $\eta=1$, and
$\Omega=0.01U$.}\label{fig7}
\end{figure}

At the quantum critical point, for $\delta=0$,
$\lambda_-(\omega=0,\pm k_x=\pm k_y=k_0)$ touches zero which
signifies the destabilization of the MI phase. The expression for
$k_0$ and the critical values of $t$ and $\gamma$ at which this
happens can be found from the conditions $\lambda_-(\omega=0,\pm
k_x=\pm k_y=k_0)=0$ and $\partial \lambda_-(\omega=0,\pm k_x=\pm
k_y=k_0)/\partial k_{0}=0$ and yields (with $F_{\pm} \equiv
F_{\pm}(\omega=0)$)
\begin{eqnarray}
F_+ + 8t \cos(k_0) &=& \sqrt{F_-^2 + 32 \gamma^2 \sin^2(k_0)}
\nonumber\\
\sin(k_0) \left(\cos(k_0) + \frac{F_+ t}{4 \gamma^2 + 8t^2} \right)
&=& 0 \label{cond1}
\end{eqnarray}
Eqs.\ \ref{cond1} provide us the position of the critical point and
allows to find $\gamma_c/U$ ($t_c/U$) and $k_0$ for any given $t/U$
($\gamma/U$), $\mu/U$, $\lambda$, and $\Omega/U$. The basic features
of the solution to Eq.\ \ref{cond1} is as follows. For $\gamma=0$,
the only solution of Eq.\ \ref{cond1} is $k_0=0$ and $t_c^{(1)}=
(|F_-| -F_+)/8$. As we turn of a finite $\gamma$, another possible
solution emerges at $k'_0 =\arccos(F_+t_c^{(2)}(\gamma)/(4 \gamma^2
+ 8[t_c^{(2)}(\gamma)]^2)$ where $t_c^{(2)}$ is the solution of $F_+
+ 8t_c^{(2)}(\gamma) \cos(k'_0) = \sqrt{F_-^2 + 32 \gamma^2
\sin^2(k'_0)}$. Depending on the chosen $\mu$, $\Omega$, and
$\lambda$, there is a critical value of $\gamma= \gamma_{0c}$, at
which $t_c^{(2)}(\gamma) \le t_c^{1}$. At this value of
$\gamma_{0c}$, $k_0$ shifts to a non-zero value $k'_0$. A similar
behavior may be inferred by choosing a fixed $t$ and by slowing
increasing $\gamma$ to reach the transition. In particular we note
that in such cases, for $t=0$, $k_0= \pi/2$ and $\gamma_c =
\sqrt{F_1 F_2/8}$. A plot of the phase-diagram based on Eq.\
\ref{cond1} is shown in Fig.\ \ref{fig6}. The top left panel of
Fig.\ \ref{fig6} shows the MI-SF phase diagram in the $\mu-\gamma$
plane for specific $t$ while the top right panel exhibit the phase
diagram in the $t-\mu$ plane for specific $\gamma$. These plots are
qualitatively similar to their mean-field counterparts in Fig.\
\ref{fig5} and exhibit reentrant SI transition as a function of
$\gamma$ for any non-zero $t$. The bottom panels of Fig.\ \ref{fig6}
shows the phase diagrams for finite $\delta$ (computed numerically
starting from the expression of $G_0$ for finite $\delta$ in Eq.\
\ref{gfmott} and using the method outlined in this section) which
are seen to be qualitatively similar to their $\delta=0$
counterparts. The left panel of Fig.\ \ref{fig7} shows a plot of
$k_0$ as a function of $t_c(\mu)$ for several representative values
of $\gamma$ with $\lambda=0.4$, and $\Omega=0.01 U$. We find that
for small $\gamma$, there is a finite range of $t_c$ for which the
transition takes place at $k_0=0$. The width of this region shrinks
with increasing $\gamma$ and beyond a critical $\gamma=\gamma'_c$,
the transition always takes place with finite $k_0$. For
$\lambda=0.4$ and $\Omega=0.01 U$, we find $\gamma'_c \simeq 0.033
U$ as can be seen from the left panel of Fig.\ \ref{fig7}.

The critical theory for the MI-SF transition can now be constructed
in terms of the low-energy excitations around $\omega=0$ and $k=k_0$
which can be described by a set of bosonic fields $\varphi_i(k)$
around each of these minimum. For $n$ such minima at $k_0 \equiv
k_0^i$ where $i=1..n$, one expresses the field $\Phi_- =
\Phi_-(\omega=0,{\bf k}=k_0^i) +\varphi_i(k)$ and obtain the
quadratic action
\begin{eqnarray}
S_2^{\rm cr} &=& -\int \frac{d^2k d\omega}{(2\pi)^3} \sum_{i=1}^n
\varphi_i^{\ast} (k) \left( K_0 \omega + K_1 \omega^2 -v^2 |{\bf
k}|^2
\right) \varphi_{i} (k) \nonumber\\
K_0 &=& \frac{\partial \lambda_-}{\partial \omega}\Big|_{\omega=0,
{\bf k}=k_0^i} = \frac{1}{2} \Big[ F'_+(0)
-\frac{F_-(0)F'_-(0)}{F_+(0)+8t \cos(k_0^i)} \Big]
\nonumber\\
K_1 &=& \frac{1}{2} \frac{\partial^2 \lambda_-}{\partial
\omega^2}\Big|_{\omega=0, {\bf k}=k_0^i} = \frac{1}{4}\Big[F^{''}_+(0) \nonumber\\
&& -\frac{F_-^{'2}(0) + F^{''}_-(0)F_-(0)}{F_+(0)+8t\cos(k_0^i)}
+\frac{
[F'_-(0)F_-(0)]^2}{[F_+(0)+8t\cos(k_0^i)]^3} \Big] \nonumber\\
v^2 &=& \frac{1}{2} \frac{\partial^2 \lambda_-}{\partial
(k_0^{i})^2}\Big|_{\omega=0, {\bf k}=k_0^i}  = -2 \Big[ t
\cos(k_0^i)\nonumber\\
&& +\frac{8 \gamma^2 \cos(2k_0^i)}{F_+(0)+8t \cos(k_0^i)} + \frac{
64 \gamma^4 \sin^2(2k_0^i)}{[F_+(0) + 8t\cos(k_0^i)]^3}\Big]
\label{quadcoeff}
\end{eqnarray}
where $'$ denotes differentiation with respect to $\omega$. From
Eq.\ \ref{quadcoeff}, we find that the critical theory has dynamical
critical exponent $z=2$ except for special points at which $K_0=0$
leading to $z=1$. In usual MI-SF transition this point appears to be
at the tip of the MI lobe. Here we find a line of such $z=1$
transitions in the $t-\gamma$ plane as shown in the right panel of
Fig.\ \ref{fig7} for representative values $\delta=0$,
$\lambda=0.4$, $\eta=1$, and $\Omega=0.01 U$.

The structure of the quadratic part of the critical action found in
Eq.\ \ref{quadcoeff} remains qualitatively similar for $\delta \ne
0$ except for two differences. The first difference in the effective
action comes from the fact that the number of minima is halved due
to the lifting of $k_x \to -k_x$ symmetry as discussed in Sec.\
\ref{miphaseb} while the second difference stems from the fact that
$v_x \ne v_y$ for $\delta \ne 0$ leading to an anisotropic
dispersion of the critical theory. Consequently, the critical action
$S_2$ now has the form
\begin{eqnarray}
S_2^{\rm cr; \delta \ne 0} &=& - \int \frac{d^2k d\omega}{(2\pi)^3}
\sum_{i=1}^n \varphi_i^{\ast} (k) \left( K_0 \omega + K_1 \omega^2
\right. \nonumber\\
&& \left.-v_x^2 k_x^2 -v_y^2 k_y^2 \right) \varphi_{i} (k)
\label{quadcoeff2}
\end{eqnarray}
The positions of the $z=1$ line in the $\mu-\gamma$ plane also
changes. However, rest of the features remain the same. In Sec.\
\ref{crth}, we shall analyze the critical theory in details and show
that the MI-SF transition at $t/\gamma=0$ is unconventional in the
sense that it is accompanied by the emergence of an additional
gapless mode at criticality.

\subsection{Analysis of the critical theory}
\label{crth}

Having established the analytical form for $S_2^{\rm cr}$ for
$\delta=0$, we shall now analyze the effective Landau-Ginzburg
theory for the transition. The quadratic part of the effective
action remains the same as in Eq.\ \ref{quadcoeff}. Our analysis
shall hold for $\delta \ne 0$ as well; in this case Eq.\
\ref{quadcoeff} shall be substituted by Eq.\ \ref{quadcoeff2}. In
this section, we shall not bother with microscopic calculation;
instead we shall analyze the critical theory from the symmetry
perspective as done, for example, in Ref.\ \onlinecite{balents1},
for small $t/\gamma$ where the minima of $G^{-1}(k)$ occurs at
non-zero ${\bf k}_0^{\pm}= (k_0, \pm k_0)$. In the presence of two
such minima, the bosonic field can be written as
\begin{eqnarray}
\Phi_-({\bf r},t )= \varphi_1({\bf r},t) e^{i{\bf k_0^+}\cdot {\bf
r}} + \varphi_2 ({\bf r},t) e^{i{\bf k_0^-}\cdot {\bf r}}.
\label{chiphi}
\end{eqnarray}
Substitution of Eq.\ \ref{chiphi} in Eq.\ \ref{gdiagd0} and
subsequent expansion in $\omega$ and ${\bf k}$ (around ${\bf
k}_0^{\pm}$) leads to Eq.\ \ref{quadcoeff} with $n=2$.

To obtain the quartic action, one can in principle substitute Eq.\
\ref{chiphi} in Eq.\ \ref{quarac}, average over the fast oscillating
components involving various powers of $\cos({\bf k}_0^{\pm} \cdot
{\bf r})$ and $\sin({\bf k}_0^{\pm} \cdot {\bf r})$ which appears in
the expression of $S_4$, and obtain an effective critical action in
terms of $\varphi_{1,2}$. However, such an averaging proves to be
tricky when $\pi /(k_0^{\pm} a)$ do not turn out to be a small
integer since one may have to sum over an arbitrary large number of
lattice sites for achieving a proper averaging \cite{sinha1}. Also,
for irrational $k_0^{\pm}$, such an averaging procedure is
ill-defined. For our case, since $k_0^{\pm}$ is a continuous
function of $t/\gamma$, we adopt a symmetry-based general method for
deriving the fourth order term in the action.

The symmetry based derivation of the effective action relies on the
fact that an effective low-energy Landau-Ginzburg action describing
a phase transition must be invariant under the projective symmetry
group (PSG) transformation of its underlying lattice
\cite{balents1}. The elements of PSG for a square lattice are
translation by a lattice vector along $x$ and $y$ ($T_x$ and $T_y$),
rotation about the $z$ axis by $\pi/2$ ($R_{\pi/2}$), and reflection
about $x$ and $y$ axes ($P_x$ and $P_y$). Following the method
derived in Ref.\ \onlinecite{balents1} and using Eq.\ \ref{chiphi},
we find that under these transformation the bosonic field
$\varphi({\bf r},t)$ transforms as
\begin{eqnarray}
&& T_x: \varphi_1 \rightarrow  e^{i k_0  a}\varphi_1, \quad
\varphi_2 \rightarrow e^{-i k_0 a}\varphi_2, \nonumber\\
&& T_y: \varphi_1 \rightarrow e^{i k_0 a}\varphi_1,~~~~ \varphi_2
\rightarrow e^{i k_0 a}\varphi_2, \nonumber\\
&& R_{\pi/2} \varphi_1 \rightarrow e^{2 i k_0 y}\varphi_1,~~~~
\varphi_2 \rightarrow e^{2 i k_0 x}\varphi_2 \nonumber\\
&& P_x: \varphi_1 \to \varphi_2, \quad \varphi_2 \to \varphi_1
\nonumber\\
&& P_y: \varphi_1 \to \varphi_2^{\ast}, \quad \varphi_2 \to
\varphi_1^{\ast} \label{psg1}
\end{eqnarray}

To find the fourth order effective action consistent with Eq.\
\ref{psg1}, we first consider the case $t=0$ for which $k_0=\pi/2$.
In this case, the most general form of the quartic action is
\begin{eqnarray}
S_4^{\rm cr} &=& \frac{g}{2} \int d^2 r dt \left[\left(
|\varphi_1|^2 + |\varphi_2|^2\right)^2 + \eta_0 (\varphi_1^{\ast}
\varphi_2 + {\rm h.c})^2 \right] \nonumber\\ \label{act0}
\end{eqnarray}
where $\eta_0$ is a constant whose value will be determined later.
Redefining the fields $ \xi_{1(2)} = (\varphi_1 +(-)
\varphi_2)/\sqrt{2}$, one gets
\begin{eqnarray}
S_4^{\rm cr} &=& \frac{g}{2} \int d^2 r dt \left[\left( |\xi_1|^2 +
|\xi_2|^2\right)^2 + \eta_0 (|\xi_1|^2-|\xi_2|^2)^2 \right] \nonumber\\
\label{act1}
\end{eqnarray}
For $\eta_0 >0$, the ground state of $S_4^{\rm cr}$ and $S_2^{\rm
cr}$ thus correspond to condensation of both the fields: $\xi_1=
\xi_1^0$ and $\xi_2= \xi_2^0 \exp(i \mu_0)$. However, the relative
phase $\mu_0$ between these two fields is not fixed by the $S_4^{\rm
cr}$. Indeed, if we construct the eighth order term $S_8$ in the
effective action, it will have a PSG allowed term $S_8 = \lambda'
\int d^2 r dt (\xi_1^{\ast} \xi_2 + {\rm h.c})^4$ which will fix
$\mu_0=m\pi/2$ for $\lambda' < 0$ and $ \mu_0 = (m+1/2)\pi/2$ for
$\lambda' > 0$ where $m$ is an integer. Thus the effective phase
mode characterized by the fluctuation of the relative phase $\mu_0$
is massive in the SF phase but is expected to become gapless when
$\lambda' \to 0$ due to irrelevance of $S_8$ at criticality.
Consequently, we expect all transitions with $\eta_0 >0$ to have an
additional gapless mode in the critical region. To compute the value
of $\eta_0$, we note that since $\pi/(k_0 a)=2$, it is possible to
compute the effective action $S_4^{\rm cr}$ by direct substitution
of Eq.\ \ref{chiphi} in Eq.\ \ref{quarac}, followed by averaging
over fast oscillating terms as shown in Ref.\ \onlinecite{sinha1}.
This procedure yields Eq.\ \ref{act0} with $\eta_0=1$. Thus we find
that for $t/\gamma=0$, the two-species bosons with spin-orbit
coupling undergoes an unconventional phase transition at
$\gamma=\gamma_c$ which are accompanied by emergence of an
additional gapless mode at the transition \cite{balents1}. We note
that this also implies that the vortices corresponding to any one of
these fields $\varphi_1$ or $\varphi_2$ will have a fractional
vorticity in the sense that a boson wavefunction would pick up a
phase $\pi$ when moved around such a vortex \cite{balents1}.
However, generating such vortices experimentally in present systems
may turn out to be difficult.

For $t/\gamma \ne 0$ where $k_0 \ne \pi/2$, we find that the only
form of the effective action which is consistent with the PSG
transformation has the form
\begin{eqnarray}
S_4^{' \rm cr} &=& \frac{g}{2} \int d^2 r dt \left[\left(
|\varphi_1|^2 + |\varphi_2|^2\right)^2 + \eta_0' |\varphi_1|^2
|\varphi_2|^2 \right] \nonumber\\ \label{act2}
\end{eqnarray}
The value of $\eta'_0$ is difficult to determine for arbitrary
$k_0$; however, for certain values of $k_0$ which satisfies
$k_0/(\pi a) \in Z$, one can determine $\eta'_0$. In all such case
we find $\eta'_0 \ge 0$. This indicates that for all $t/\gamma$,
only one of the fields $\varphi_1$ or $\varphi_2$ condenses. Thus
the MI-SF critical points for such finite $t/\gamma$ are
conventional.

\section{Mean-field Analysis}
\label{mfta}

In this section, we use a Gutzwiller wavefunction to obtain the
mean-field SI phase boundary for the system. The Gutzwiller
variational wavefunction which we shall use is given by
\begin{eqnarray}
|\psi_{i}\rangle &=& a_{i}  |1,0 \rangle_{i}  + b_{i}
|0,1\rangle_{i}
+ c_{i} |1,1 \rangle_{i}+ d_{i} |2,0 \rangle_{i} \nonumber\\
&& + e_{i}  |0,2 \rangle_{i} + f_{i} |0,0 \rangle_{i},\nonumber\\
|\Psi_G\rangle &=& \prod_{i} |\psi_{i} \rangle. \label{varwav1}
\end{eqnarray}
Note that for the purpose of charting out the phase diagram and for
describing the SF phase near the SI transition point, it is not
necessary to incorporate the higher number states since we expect
these states to have very small overlap with the ground state of the
system as can be checked by explicit numerical calculation. The
variational energy of the system can be easily computed using Eqs.\
\ref{ham0}, \ref{ham1} and \ref{varwav1} and yields
\begin{eqnarray}
E &=& \langle \Psi_G|({\mathcal H}_0 +{\mathcal H}_1|\Psi_G\rangle
= E_0 + E_1 + E_2 \nonumber\\
E_0 &=& \sum_{i} -(\mu + \Omega) |a_{i}|^2 -(\mu - \Omega)|b_{i}|^2
+ (\lambda-2\mu)
|c_{i}|^2 \nonumber\\
&& + (1-2\mu-2\Omega)|d_{i}|^2 + (1-2\mu+2\Omega) |e_{i}|^2 \nonumber\\
&& + i \delta [b_{i}a^*_{i}-a_{i} b^*_{i}
+ \sqrt{2}c_{i}d^*_{i} -\sqrt{2}c^*_{i} d_{i} \nonumber\\
&& + \sqrt{2}c^*_{i}e_{i}-\sqrt{2}c_{i} e^*_{i}], \nonumber \\
E_1 &=& -t_1 \sum_{\langle {i j}\rangle} (\Delta_{i 1}^{\ast}
\Delta_{{j} 1}, + \eta  \Delta_{i
2}^{\ast} \Delta_{j 2}) + {\rm h.c.} \nonumber \\
E_{2} &=& -\gamma \Big[ \sum_{\langle {i j_x }\rangle} (\Delta_{i
1}^{\ast} \Delta_{j_x 2} - \Delta_{i 2}^{\ast}
\Delta_{j_x 1} ) \nonumber\\
&& + i \sum_{\langle i j_y \rangle} (\Delta_{i 1}^{\ast} \Delta_{j_y
2} + \Delta_{i 2}^{\ast} \Delta_{j_y 1} ) \Big] + {\rm h.c.}
\label{mften}
\end{eqnarray}
where $\sum_{\langle i j\rangle}$ denotes sum over both $x$ and $y$
neighbors of site ${i}$ while $\sum_{\langle i j_{x(y)}\rangle}$
denotes sum over $x(y)$ neighboring sites of ${i}$, and the order
parameter $\Delta_{{i} a} = \langle b_{{i} a} \rangle$ can be
expressed in terms of the Gutzwiller wavefunction coefficients as
\begin{eqnarray}
\Delta_{{i} 1} &=& f_{i}^{\ast} a_{i} + b_{i}^{\ast} c_{i} +
\sqrt{2} a_{i}^{\ast} d_{i}
\nonumber\\
\Delta_{{i} 2} &=& f_{i}^{\ast} b_{i} + a_{i}^{\ast} c_{i} +
\sqrt{2} b_{i}^{\ast} e_{i} \label{op1}
\end{eqnarray}

\begin{figure}
\includegraphics[width=\linewidth]{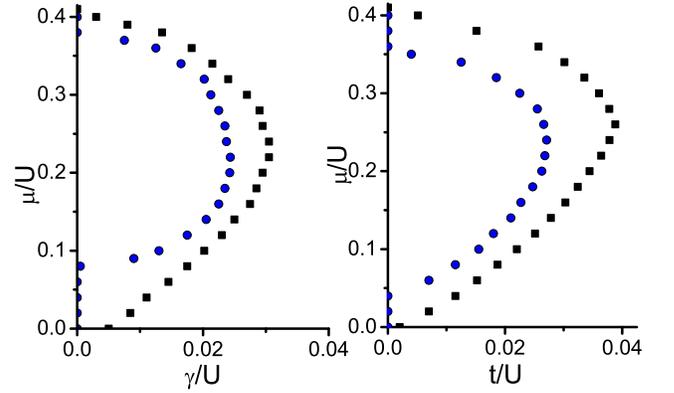}
\caption{Left panel: The MI-SF phase boundary from mean-field theory
for different representative values of $t/U=0.0$ (black squares),
$t/U=0.02$ (blue circles) in the $\mu-\gamma$ plane. Right panel:
The phase boundary in the $\mu-t$ plane for $\gamma/U=0$(black
squares) and $\gamma/U=0.02$ (blue circles). Here we have taken
$\eta=0.5$, $\delta=0.0$ and $\Omega=0.01 U$.} \label{fig5}
\end{figure}

The phase diagram obtained by numerical minimization of Eq.\
\ref{mften} is shown in Fig. \ref{fig5}. We note that for these
bosons, SI transition can be induced either by tuning $\gamma$ or
$t$. We first consider the case of $\delta=0$, $\eta=0.5$ and
$\Omega=0.01 U$. The MI phase for the parameter values is
characterized by $n_1=1$ and $n_2=0$. The MI-SF phase diagram, in
the $\mu-\gamma$ plane, is shown in the left panel of Fig.\
\ref{fig5} for representative values of $t/U=0,\,0.02$. Here, we
find that for all values of $\mu$, the transition always takes place
into a SF phase with $\langle b_1\rangle, \langle b_2 \rangle \ne
0$. Following the nomenclature of Ref.\ \onlinecite{issacson1}, we
term this SF phase as 2-SF. We also note that for any finite $t$,
the bosons display reentrant SI transition with variation of
strength of $\gamma$. The MI-SF phase diagram in the $\mu-t$ plane
for representative values of $\gamma=0,\,0.02$ is shown in the right
panel of Fig.\ \ref{fig5}. Here for $\gamma \ne 0$, we find that the
transition always takes into a 2-SF phase. In contrast, for
$\gamma=0$, a small region in the phase diagram near $\mu=0$ exhibit
1-SF superfluidity for which $\langle b_1 \rangle \ne 0$ and
$\langle b_2 \rangle =0$. The phase diagram with small non-zero
$\delta$ turns out to be qualitatively similar.

The most striking point about the superfluid phase into which the
transition takes place becomes evident on examining the values of
$\Delta_{{i}, a}$ for the ground state configuration in the SF
phase. We find that although the amplitudes of the superfluid order
parameters remain homogeneous, their phases vary with positions for
finite $\gamma/t$; in other words, the superfluid ground state
realized is an example of a twisted superfluid phase
\cite{comment1}. We also note that the relative phases between the
$x$ and the $y$ neighboring links are different leading to an
anisotropic twist. To obtain an qualitative understanding of the
role of spin-orbit coupling in the realization of such a twisted
superfluid phase, we note that these phases contribute to the energy
of the system through the terms $E_1$ and $E_2$ in Eq.\ \ref{mften}.
Taking cue from the numerical result that the magnitude of the order
parameters remain constant in the ground state configuration, we now
write $\Delta_{{i} a} = \Delta_{0a} \exp(i \phi_{{i} a})$. In what
follows, we choose the phase of the order parameter on the ${\rm
i^{th}}$ and the neighboring sites as
\begin{eqnarray}
\phi_{i1} &=& 0, \quad \phi_{i2} = \phi_0
\nonumber\\
\phi_{j_{\alpha}1} &=& \alpha_{j_{\alpha}}, \quad \phi_{j_{\alpha}
2} = \beta_{j_{\alpha}}, \label{phasedefa1}
\end{eqnarray}
where the subscript $\alpha$ takes values $x$ and $y$. Using this,
one can write $E_1$ and $E_2$ in terms of the relative phases
between the $x$ and $y$ neighbors
\begin{eqnarray}
\frac{E_1}{t \Delta_{0 1}^2} &=& -\sum_{\langle ij \rangle}
\Big\{[\cos(\alpha_{j_x}) + \cos(\alpha_{j_y}) ] \nonumber\\
&& + \eta \kappa^2 [\cos(\beta_{j_x}-\phi_0) +
\cos(\beta_{j_y}-\phi_0)] \Big \} \nonumber\\
\frac{E_2}{t \Delta_{0 1}^2} &=& - \frac{\gamma \kappa}{t} \Big[
\sum_{\langle ij_x \rangle} \cos(\beta_{j_x})
+ \cos(\alpha_{j_x}-\phi_0) \nonumber\\
&& + \sum_{\langle ij_y\rangle} \sin(\beta_{j_y}) + \sin(
\alpha_{j_y}-\phi_0)\Big]  \label{cosen}
\end{eqnarray}
where $\kappa= \Delta_{02}/ \Delta_{01}$.

\begin{figure}
\includegraphics[width=\linewidth]{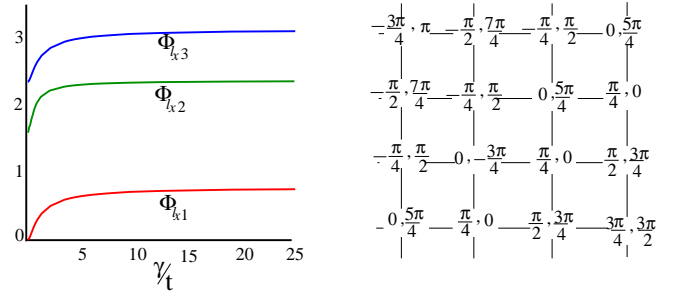}
\caption{Left panel: Plot of the relative phases $\Phi_{\ell_{x,y}
1,2,3}$ on the links of the square lattice as a function of
$\gamma/t$ for $\eta=0.5$, $\Omega=0.01 U$. Right panel:
Distribution of the phases $\phi_{{i} a}$ for the superfluid order
parameter for $\gamma/t \gg 1$.} \label{figphase}
\end{figure}

Next, we define relative phases living on $x$ and $y$ links of the
2D square lattice as
\begin{eqnarray}
\Phi_{\ell_x 1}&=&\phi_{j_{x}1}, \quad \Phi_{\ell_y 1}= \phi_{j_{y}1}\nonumber\\
\Phi_{\ell_x 2}&=&\phi_{j_{x}2}-\phi_0, \quad \Phi_{\ell_y 2}= \phi_{j_{y}2}-\phi_0\nonumber\\
&&\Phi_{\ell_x 3}=\phi_{j_{x}1}-\phi_0.
\end{eqnarray}
In terms of these phases, Eq.\ \ref{cosen} can be recast as
\begin{eqnarray}
\frac{E_1}{t \Delta_{0 1}^2} &=& -\sum_{\ell_x,\ell_y}\Big\{
[\cos(\Phi_{\ell_x 1}) + \cos(\Phi_{\ell_y 1}) ] \nonumber\\
&& + \eta \kappa^2 [\cos(\Phi_{\ell_x 2}) +
\cos( \Phi_{\ell_y 2})]\Big\}\nonumber\\
\frac{E_2}{t \Delta_{0 1}^2} &=& - \frac{\gamma \kappa}{t}
\sum_{\ell_x,\ell_y} \Big [\cos(\Phi_{\ell_x 3})
- \cos(\Phi_{\ell_x 2}-\Phi_{\ell_x 3}+\Phi_{\ell_x 1}) \nonumber\\
&& + \sin(\Phi_{\ell_x 2}-\Phi_{\ell_x 3}+\Phi_{\ell_y 1})\nonumber\\
&&+ \sin( \Phi_{\ell_y 2}-\Phi_{\ell_x 2}+\Phi_{\ell_x 3}) \Big]
\label{cosen1}
\end{eqnarray}
From Eq.\ \ref{cosen1}, we clearly see that unless $\gamma/t$ is
small, the minimal energy configuration correspond to non-zero but
uniform values relative phases over $x$ and $y$ links. Note that the
precise numerical values of these phases depend on $\kappa$ and
hence requires input from numerical minimization of Eq.\
\ref{mften}. However, once we know the value of $\kappa$, we find
that the relative phases for the minimum energy are the solutions of
the coupled transcendental equations $\partial (E_1+E_2)/\partial
\Phi_{\ell_{x(y)} 1(2, 3)}= 0$ which yields
\begin{eqnarray}
&& \sin(\Phi_{\ell_x 1})- \frac{\gamma \kappa}{t}\sin(\Phi_{\ell_x 2}-
\Phi_{\ell_x 3}+\Phi_{\ell_x 1})=0\nonumber\\
&&\sin(\Phi_{\ell_y 1})-\frac{\gamma \kappa}{t}\cos(\Phi_{\ell_x 2}
-\Phi_{\ell_x 3}+\Phi_{\ell_y 1})=0\nonumber\\
&&\eta\kappa^2\sin(\Phi_{\ell_x 2})- \frac{\gamma \kappa}{t}
\Big\{\sin(\Phi_{\ell_x 2}-\Phi_{\ell_x 3}+\Phi_{\ell_x 1})+\cos(\nonumber\\
&&\Phi_{\ell_x 2}-\Phi_{\ell_x 3}+\Phi_{\ell_y 1})
-\cos(\Phi_{\ell_y 2}-\Phi_{\ell_x 2}+\Phi_{\ell_x 3})\Big\}=0\nonumber\\
&&\eta\kappa^2\sin(\Phi_{\ell_y 2})- \frac{\gamma \kappa}{t}
\sin(\Phi_{\ell_y 2}-\Phi_{\ell_x 2}+\Phi_{\ell_x 3})=0\nonumber\\
&&\sin(\Phi_{\ell_x 3})+\sin(\Phi_{\ell_x 2}-\Phi_{\ell_x 3}
+\Phi_{\ell_x 1})+\cos(\Phi_{\ell_x 2}\nonumber\\
&&-\Phi_{\ell_x 3}+\Phi_{\ell_y 1})-
\cos(\Phi_{\ell_y 2}-\Phi_{\ell_x 2}+\Phi_{\ell_x 3})=0
\label{mftphase}
\end{eqnarray}
In general, these equations need to be solved numerically and we
have not found analytic solutions for them for arbitrary values of
$\kappa$ and $\gamma/t$. However, in the special case $\gamma/t \gg
1$, we find that these equations admit an easy solution
\begin{eqnarray}
\Phi_{\ell_x 1}&=&- \Phi_{\ell_y 1}=\pi/4, \quad \Phi_{\ell_x 3}=\pi
\nonumber\\
\Phi_{\ell_x 2}&=& -\Phi_{\ell_y 2}=3\pi/4\label{phasev1}
\end{eqnarray}
The corresponding phase distribution of the superfluid order
parameter $\phi_{{i} a} \equiv (\phi_{{i} 1}, \phi_{{i} 2})$ is
shown in the right panel of Fig.\ \ref{figphase}. For all values of
$\gamma/t$ and $\kappa$, we find $\Phi_{\ell_x 1(2)} = -
\Phi_{\ell_y 1(2)}$. Also, for $\eta < 1$ (which implies $\kappa
<1$), we find that $\Phi_{\ell_{x,y} 2}$ have a discontinuous jump
to finite value around $\gamma=0$. The occurrence of this can be
easily understood a competition between second and the third set of
terms ( those proportional to $\eta \kappa^2$ and $\gamma \kappa/t$
respectively) in Eq.\ \ref{cosen1}. A plot of the relative phases on
the $x$ and $y$ links is shown in the left panel of Fig.\
\ref{figphase} as a function $\gamma/t$. We find that the relative
phases take finite value for non-zero $\gamma$ and approaches those
given by Eq.\ \ref{phasev1} with increasing $\gamma/t$ thus leading
to the realization of a twisted superfluid ground state. We have
checked that the value of the relative phases obtained from
minimization of Eq.\ \ref{cosen1} agrees to those computed from
minimization of Eq.\ \ref{mften}.

\section{Collective modes}
\label{goldstone}

In this section, we use the critical theory developed in Sec.\
\ref{trsc} to obtain the collective modes in the superfluid phase
near the critical point. We first consider the case $\delta=0$ for
which one can obtain straightforward analytical expressions for
these modes. We first consider $t=0$. We begin with the quadratic
and quartic parts of the boson action in the SF phase near the
critical point which are given by
\begin{eqnarray}
S^{'}&=& S^{'}_2+ S^{'}_4 \nonumber \\
S^{'}_2&=& -\int \frac{d^2 k d \omega}{(2 \pi)^3} \sum_{i=1,2}
\xi_i^{\ast} \left(K_0 \omega + K_1 \omega^2 \right. \nonumber\\
&& \left. -v^2 |{\bf k}|^2 + |r_0| \right) \xi_i \label{scr41} \\
S^{'}_4 &=& \frac{g}{2} \int d^2 r dt \left[ (|\xi_1|^2 +
|\xi_2|^2)^2 + \eta_0(|\xi_1|^2- |\xi_2|^2)^2 \right] \nonumber
\end{eqnarray}
In the SF phase, both the fields condense with amplitudes
$|\xi_{i0}|=\sqrt{|r_0|/2g}$ for $i=1,2$. To obtain the collective
modes, we therefore expand the fields $\xi_i =\xi_{i0} + \delta
\xi_i$, where $\delta \xi_i$ represents small amplitudes fluctuating
fields which describes the collective modes of the condensate. Using
Eq.\ \ref{scr41}, we obtain an effective quadratic action for
$\delta \chi_i$. It turns out that for $\eta_0=1$, the quadratic
actions for $\delta \xi_1$ and $\delta \xi_2$ reduces to
block-diagonal form which can be written as
\begin{eqnarray}
S' &=& \int \frac{d^2 k d \omega}{(2 \pi)^3} \sum_{i=1,2} \Psi_i^{*}
{\tilde \Lambda} \Psi_i,
\nonumber\\
{\tilde \Lambda} &=& \left( \begin{array}{cc} D_0(\omega, {\bf k})
+ |r_0| & \frac{|r_0|}{2}\\
\frac{|r_0|}{2} & D_0(-\omega, -{\bf k}) +|r_0|\end{array}\right)
\label{stzero1}
\end{eqnarray}
where $\Psi_i=\left( \delta \xi^*_i, \delta \xi_i \right)^T$, and
$D_0(\omega, {\bf k}) = -(K_1 \omega^2 + K_0 \omega - v^2 |{\bf
k}|^2 + |r_0|)/2$. The collective modes corresponding to the field
$\xi_i$ can then be obtained from the condition ${\rm Det} {\tilde
\Lambda} =0$ and yields,
\begin{eqnarray}
\omega_{1(2)}({\bf k}) &=& \sqrt{\frac{ \pm \alpha_{\bf k} + \sqrt{
\alpha_{\bf k}^2 + 4 (A_{\bf k}^2-|r_0|^2) K_1^2}}{2 K_1^2} }
\nonumber\\
\alpha_{\bf k} &=& 2 K_1 A_{\bf k} + K_0^2, \quad A_{\bf k} = v^2
|{\bf k}|^2 + |r_0| \label{coll1}
\end{eqnarray}
Each of these two modes are doubly degenerate. It is easy to see
from Eq.\ \ref{coll1} that $\omega_{2}({\bf k})$ are gapped while
$\omega_1({\bf k})$ is gapless with $\omega_1 \sim |{\bf k}|^2$ at
small $|{\bf k}|$ for $K_0 \ne 0$ and $\omega_1 = |{\bf k}|
v/\sqrt{K_1}$ for $K_0=0$. The mass of the gapped mode can be read
off from Eq.\ \ref{coll1} and are given by
\begin{eqnarray}
m &=& \sqrt{ (2 |r_0| K_1 + K_0^2)/K_1^2} \label{mass1}
\end{eqnarray}
Note that in this case, there is one gapless and one gapped mode and
each of these are doubly degenerate. This leads to two gapless modes
in the SF phase which is a consequence of condensation of both
$\xi_1$ and $\xi_2$ at the transition.

\begin{figure}
\includegraphics[width=\linewidth]{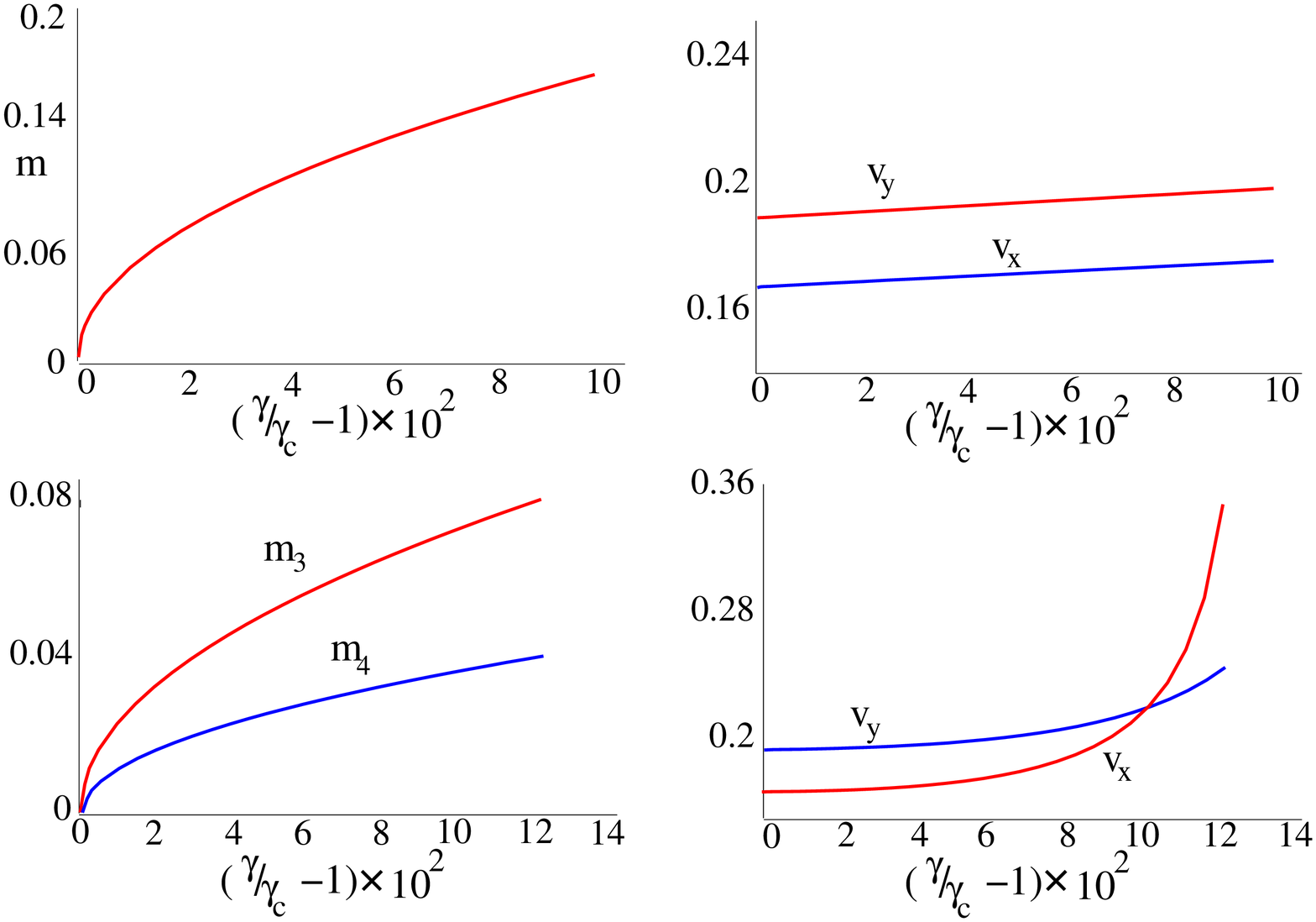}
\caption{Left upper panel: Plot of the mass of the gapped mode for
$t=0$ and $\delta=0.005 U$ as a function of $(\gamma/\gamma_c-1)$.
Upper right panel: Plot of the velocities of the gapless modes
($v_x$ and $v_y$) as a function of $(\gamma/\gamma_c-1)$. Lower left
(right) panel: Mass of the non-condensed modes (Velocity of the
gapless mode) for $ t/\gamma=0.005$, $K_0=0$ and $\delta=0.005
U$.}\label{fig8}
\end{figure}

Next we consider the case $t/\gamma \ne 0$. In this case, we begin
with the action
\begin{eqnarray}
S^{''} &=& S^{''}_2+ S^{''}_4 \nonumber \\
S^{'}_2 &=& - \int d^2r dt \sum_{i=1,2} \varphi_i^{\ast}
\left(K_0 \omega + K_1 \omega^2 -v^2 |{\bf k}|^2 + |r_0| \right) \varphi_i \nonumber \\
&&S^{''}_4= \frac{g}{2} \int d^2 r dt \left[ (|\varphi_1|^2 +
|\varphi_2|^2)^2 +\eta'_0 |\varphi_1|^2 |\varphi_2|^2\right]. \nonumber\\
\label{scr42}
\end{eqnarray}
To obtain the collective modes, we note that the field $\varphi_1$
condenses with an amplitude $|\varphi_{01}| = \sqrt{|r_0|/g}$. We
then expand the fields $\varphi_1 = \varphi_{01} + \delta \varphi_1$
and $\varphi_2 = \delta \varphi_2$ and obtain the effective
quadratic action for the field $\delta \varphi_{1,2}$. It turns out
that these actions decouple. The effective action for $\delta
\varphi_1$ turns out to be analogous to Eq.\ \ref{stzero1} and
yields a gapless and a gapped mode with $\omega= \omega_{1(2)}({\bf
k})$. The effective action for $\delta \varphi_2$ is given by
\begin{widetext}
\begin{eqnarray}
S^{''} &=&\int \frac{d^2 k d \omega}{(2 \pi)^3}  \Psi_2^{'*}
\Lambda^{''} \Psi'_2,
\nonumber\\
\Lambda^{''} &=& \left( \begin{array}{cc} D_0(\omega, {\bf k}) + (2+\eta'_0) |r_0|/2 & 0 \\
0 & D_0(-\omega, -{\bf k}) +(2+\eta'_0)|r_0|/2 \end{array}\right)
\label{stzero2}
\end{eqnarray}
\end{widetext}
where $\Psi'_2 =[\varphi_2 (\omega,{\bf k}),
\varphi_2^{\ast}(-\omega,-{\bf k})]^T$. The collective modes
obtained using Eq.\ \ref{stzero2} are given by
\begin{eqnarray}
\omega_{3(4)}({\bf k}) &=& \frac{ -(+) K_0 + \sqrt{ K_0^2 + 4 K_1
(A_{\bf k} + \eta'_0 |r_0|)} }{2 K_1} \label{coll2}
\end{eqnarray}
The masses of these modes are given by
\begin{eqnarray}
m_{3(4)} &=& \frac{ -(+) K_0 + \sqrt{ K_0^2 + 4 K_1 |r_0| (1+
\eta'_0)} }{2 K_1}
\end{eqnarray}
Thus in this case, we have one gapless and three gapped mode. We
note that since $K_0$, $K_1$ and $v$ can be computed from
microscopic parameters of the theory, our analysis provides a way of
obtaining the velocities and masses of the gapped and the gapless
collective modes directly from the parameters of the microscopic
Hamiltonian of the bosons.

The inclusion of finite $\delta$ changes this picture in two
essential ways. First, it lifts the degeneracy between some of the
modes. Second, it makes the dispersion anisotropic since in the
presence of a finite $\delta$, $v_x$ and $v_y$ are not identical. A
plot of the masses of the gapped and velocity of the gapless modes
for a finite but small $\delta=0.005U$ is shown in Fig.\ \ref{fig8}.
In accordance with the expectation, we find that the velocities of
the gapless modes are different.

\section{Discussion}
\label{conc}

In this work, we have studied the SI transition of two-species
bosons with spin-orbit coupling. The main conclusions of our work
are the following. First we have shown, via explicit calculation of
the boson momentum distribution function, that the SI transition is
accompanied by precursor peaks in the MI phases near the transition
and that the position of these peaks can be tuned by tuning the
strength of the spin-orbit coupling. We note that this feature of
our theory can be directly verified experimentally by routine
momentum distribution measurements \cite{Greiner1,spielman1}.
Second, we have analyzed the MI-SF phase boundary and have shown the
existence of reentrant SI transitions at fixed $t$ and $\eta$ with
variation of $\gamma$. This feature can also be detected
experimentally by momentum distribution measurements. Third, we have
shown that for $t/\gamma=0$, the SI transition is unconventional in
the sense that it is accompanied by emergence of a gapless mode in
the critical region. Fourth, we have computed the collective modes
in the SF phase near the transition. We have presented analytical
formulae for the gapless and the gapped mode and have provided
explicit expression for their masses and velocities in terms of
microscopic parameters of theory. These predictions can be verified
by routine spectroscopy measurements on these systems
\cite{esslinger1}. Finally, our mean-field study has revealed the
presence of a twisted superfluid ground state in these systems with
an anisotropic twist angle whose magnitude depend on $\gamma/t$.

KS thanks DST for support through grant SR/S2/CMP-001/2009.

\end{document}